\documentclass[aps,reprint,pra,longbibliography,superscriptaddress]{revtex4-1}
\usepackage{amssymb}
\usepackage{amsmath}
\usepackage{dcolumn}
\usepackage{bm}
\usepackage{graphicx}
\usepackage{mathrsfs}
\usepackage[colorlinks,linkcolor=blue,anchorcolor=blue,citecolor=blue,urlcolor=blue]{hyperref}

\begin{document}

\title{Simulation of higher-order topological phases in 2D spin-phononic crystal networks}
\author{Xiao-Xiao Li}
\affiliation{Shaanxi Province Key Laboratory of Quantum Information and Quantum Optoelectronic Devices,
Department of Applied Physics, Xi'an Jiaotong University, Xi'an 710049, China}
\affiliation{Department of Physics, University of Oregon, Eugene, Oregon 97403, USA}
\author{Peng-Bo Li}
\email{lipengbo@mail.xjtu.edu.cn}
\affiliation{Shaanxi Province Key Laboratory of Quantum Information and Quantum Optoelectronic Devices,
Department of Applied Physics, Xi'an Jiaotong University, Xi'an 710049, China}


\begin{abstract}
We propose and analyse an efficient scheme for simulating higher-order topological phases of matter in two dimensional (2D) spin-phononic crystal networks.  We show that, through a specially designed periodic driving, one can selectively control and enhance the bipartite silicon-vacancy (SiV) center arrays, so as to obtain the chiral symmetry-protected spin-spin couplings. More importantly, the  Floquet engineering spin-spin interactions support rich quantum phases associated with topological invariants. In momentum space, we analyze and simulate the topological nontrivial properties of the one- and two-dimensional system, and show that  higher-order topological phases can be achieved under the appropriate periodic driving parameters.  As an application in quantum information processing, we study the robust quantum state transfer via topologically protected edge states. This work opens up new prospects for studying quantum acoustic, and offers an experimentally feasible platform for the study of higher-order topological phases of matter.

\end{abstract}

\maketitle
\section{introduction}

Topological insulators (TIs) possess topologically protected surface or edge states, which can be utilized as robust transmission channels. In condensed matter physics, topological systems such as the quantum Hall effect and the quantum spin Hall effect have been extensively studied \cite{hasan2010colloquium,thouless1982quantized,kane2005quantum}. With the combination of topology and quantum theory, topological protection has developed some interesting applications in quantum information processing. In photonics, topological edge states can be used to realized one-way transport without breaking time reversal symmetry \cite{lodahl2017chiral,barik2018topological,he2010tunable,peano2015topological,haldane2008possible,ozawa2019topological}. In quantum computation, topology was introduced to solve the decoherence problem, in which the non-Abelian topological phases of matter are used to encode and manipulate quantum information \cite{stern2013topological,nayak2008non,xu2020tensor}.

The Su-Schrieffer-Heeger (SSH) model, originally derived from the dimerized chain, serves as the simplest example of one-dimensional (1D) topological insulator \cite{PhysRevLett.42.1698}. So far, the SSH model have been realized in a number of quantum structures. For instance, a recent experiment demonstrated a tunable dimerized model and observed the topological magnon insulator states in a superconducting qubit chain \cite{cai2019observation}. As for ion-trap or optical lattice systems, an external periodic driving is generally needed to trigger the topological properties of the system, realizing the Floquet topological insulators in these systems \cite{PhysRevLett.119.210401,PhysRevLett.99.110501,PhysRevLett.123.126401,liu2019floquet,plekhanov2017floquet}. In addition, to investigate the topological characters of high-dimensional quantum devices, several theoretical works extended the SSH model to the two-dimensional (2D) case \cite{PhysRevA.100.032102,PhysRevApplied.12.034014,PhysRevB.100.075437,PhysRevB.100.075120,PhysRevLett.118.076803,PhysRevLett.122.233903}. However, with the present experimental conditions, the observation of topological phenomena, in particular the higher-order topology,  in the quantum domain is still challenging \cite{yan2019higher,schindler2018higher,zhang2019second}.

In recent years, quantum acoustics has aroused growing interests, which mainly studies the coherent interactions between quantized phonon modes and quantum emitters. Mechanical resonators or propagating phonons with low speed of sound in solids, offer unique advantages for transmitting quantum information between solid-state quantum systems. To date, experimental and theoretical progress has realized a variety of hybrid mechanical structures involving a large number of different quantum systems, such as solid-state defects \cite{maity2020coherent,lemonde2018phonon,li2016hybrid,li2015hybrid,li2018hybrid,bienfait2019phonon,kuzyk2018scaling,li2017preparing}, superconducting circuits \cite{etaki2008motion,satzinger2018quantum,chu2018creation,lahaye2009nanomechanical,dong2019multiphonon,li2020enhancing}, ultracold atoms \cite{camerer2011realization,jockel2015sympathetic}, and quantum dots \cite{metcalfe2010resolved,yeo2014strain}. Among these, due to the excellent coherence properties even at room temperature, defect spins in diamond and silicon carbide have become one of the most promising systems for quantum applications in solid states. In particular, the negatively charged silicon-vacancy (SiV) center in diamond serves as an emerging block for hybrid quantum systems because of high strain susceptibility and remarkable optical properties \cite{meesala2018strain,hepp2014electronic,sohn2018controlling,qiao2020phononic}.

\begin{figure*}[tbp]
\includegraphics[width=15cm]{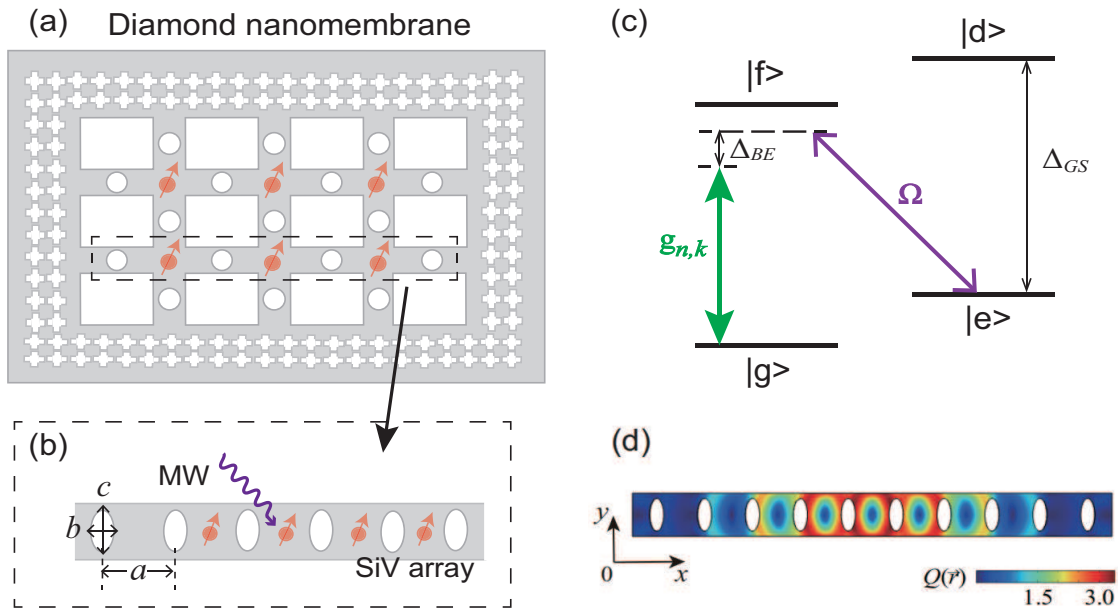}
\caption{(Color online) (a) Schematics of the hybrid device studied in this
work. Nanomechanical 2D phonon band-gap setups elaborately designed with high-Q cavities in a patterned diamond membrane. (b) Arrays of SiV color centers are implanted evenly in the phononic
waveguide. The lattice constant of the phononic crystal is $a=100$ nm, the size of the ellipse hole are $(b,c)=(30,76)$ nm, and the thickness is $t=20$ nm. SiV spins are
driven by microwave driving fields. (c) Ground-state energy levels of a single SiV center. Lower and upper states split via an external magnetic field. $%
g_{n,k}$ describes the coupling strength between the spin and the phonon with the transition $(|g\rangle \leftrightarrow|f\rangle )$, and $\Delta
_{BE}=\omega_{s}-\omega_{BE}$ is the detuning between the spin transition
and phononic band edge frequency. (d) Displacement pattern of the phononic compression
mode at the band edge frequency $\omega_{BE}$.}
\label{fig_model}
\end{figure*}

Previous works have shown that a highly tunable spin-phonon interaction can be achieved near a phononic band gap \cite{li2019band,PhysRevResearch.2.013121,lemonde2019quantum}. Phononic crystals, defined as elastic waves propagating in periodic
structures, which provide a powerful candidate for manipulating the interplay of phonons and other quantum systems. Because of the unique band structure of the phononic crystal, a single phonon bound state emerges within the band gap \cite{liu2017quantum,john1990quantum,krinner2018spontaneous}, resulting in a stronger and controllable spin-phonon coupling.
More importantly, owing to the advantage of the scalable nature of nanofabrication, the spin-phononic crystal setup is experimentally feasible when extending to the higher dimensional case \cite{chan2012optimized,burek2016diamond,kuang2004effects,zhang2004negative,pennec2010two,sukhovich2008negative,ding2019experimental,serra2018observation,he2016acoustic,safavi2010optomechanics,lemonde2019quantum,peng2019chirality}.

In this work, we propose an efficient protocol for studying the topological quantum properties in 2D SiV-phononic crystal networks. Driving the SiV color center arrays with the periodic microwave fields, we obtain the Floquet engineering spin-spin interactions with some unique properties. We find that, it is possible to selectively control the phonon band-gap mediated spin-spin couplings by modulating the parameters of the periodic driving. We show that, the chiral symmetry-protected spin-spin interactions are attained in the bipartite SiV center arrays, and more importantly, the Floquet engineering spin-spin interactions support rich quantum topological phases.
To investigate the topological nontrivial features of the system, we convert the 1D and 2D Floquet engineering spin-spin Hamiltonian to the momentum space. Firstly, we study the topological invariant Winding and Chern numbers, respectively. Apart from the original definitions, here we offer a geometrically intuitive way to calculate the topological invariant. And then we obtain the 1D and 2D topological Zak phases, respectively. We show that, the higher-order topological phase can be achieved under the appropriate periodic driving parameters. In addition, we give the analytical and numerical solutions of the topological edge states. Finally, we study the topological protected quantum state transfer and discuss the effect of SiV spin dephasing. This work offers an
experimentally feasible platform for studying topological nontrivial phenomena in higher-dimensional
quantum systems.

\section{2D spin-phononic crystal networks}

The 2D spin-phononic crystal setup is depicted in Fig.~$%
1(a)$, where identical nodes are arranged in a square lattice. The diamond
waveguide is perforated with periodic elliptical air holes, which yields the
tunable phononic band structures. SiV color centers are evenly
located at the nodes of the phononic structure, which are coupled to the
acoustic vibrations via lattice strain. The pattern structure of the edge of the
diamond membrane is designed to ensure the high-Q phonon band-gap cavities \cite{chan2012optimized}.

For the phononic crystal, we first consider a quasi-1D geometry model, which
supports acoustic guide modes $\omega _{n,k}$, with $n$ the band index and $%
k $ the wave vector along the waveguide direction. The mechanical
displacement mode profile $\vec{Q}(\vec{r},t)$ can be obtained by solving
the elastic wave equation \cite{PhysRevB.49.2313}. Analogous to the electromagnetic
field in quantum optics, the mechanical displacement field can be quantized,
i.e., $H_{p}=\sum_{n,k}\hbar \omega _{n,k}a_{n,k}^{\dagger }a_{n,k}$, with $%
a_{n,k}$ and $a_{n,k}^{\dagger }$ the annihilation and creation operators
for the phonon modes.

SiV color centers are interstitial point defects wherein a silicon atom is
positioned between two adjacent vacancies in the diamond lattice. The
negatively charged SiV center can be treated as an effective $S=1/2$ system.
For the electronic ground state of the SiV center, the $\left\vert
^{2}E_{g}\right\rangle $ states are the combination of a twofold orbital and
a twofold spin degeneracy. Considering the spin-orbit interaction and
Jahn-Teller effect, the orbital states are separated into a lower branch
(LB) and upper branch (UB) with frequency $\Delta _{GS}=2\pi \times 46$ GHz.
In the presence of an external magnetic field $\vec{B}$, the Zeeman effect
will be further split the spin degenerate states. The SiV ground state
Hamiltonian can be written as \cite{hepp2014electronic}
\begin{equation}
H_{SiV}=-\hbar \lambda _{SO}L_{z}S_{z}+H_{JT}+\hbar f\gamma
_{L}B_{z}L_{z}+\hbar \gamma _{S}\vec{B}\cdot \vec{S},
\end{equation}%
where $\lambda _{SO}$ is the strength of spin-orbit interaction, $\gamma
_{L} $ and $\gamma _{S}$ correspond to the orbital and spin gyromagnetic
ratio. Diagonalizing Eq. $(1)$, we obtain four eigenstates $\{\left\vert
g\right\rangle =\left\vert e_{-}\downarrow \right\rangle ,\left\vert
e\right\rangle =\left\vert e_{+}\uparrow \right\rangle \},\{\left\vert
f\right\rangle =\left\vert e_{+}\downarrow \right\rangle $ and $\left\vert
d\right\rangle =\left\vert e_{-}\uparrow \right\rangle \}$, where $%
\left\vert e_{\pm }\right\rangle =(\left\vert e_{x}\right\rangle \pm
i\left\vert e_{y}\right\rangle )/2$ are eigenstates of the orbital angular
momentum operator. The corresponding energy level diagram is given in Fig.~$%
1(c)$. As result, the spin-flip transitions are allowed between the four
sublevels with opposite electronic spin components \cite{PhysRevLett.120.053603,pingault2017coherent}. Specifically, the two
lowest sublevels $(|g\rangle ,|e\rangle )$ can be treated as a long-lived
qubit and coherently controlled via an optical Raman process. Furthermore,
in the high-strain limit, this transition can be directly driven with a
microwave field \cite{nguyen2019integrated}.

In the SiV-phonon system, the mechanical lattice vibration modifies the
electronic environment of the SiV center, resulting in the coupling of its
orbital states $\left\vert e_{-}\right\rangle $ and $\left\vert
e_{+}\right\rangle $. As for the setup shown in Fig.~$1(b)$, when the
transition frequency of the spin state is tuned close to the phononic band
edge, we can obtain the strong strain coupling between the SiV center and
phononic crystal mode \cite{lemonde2018phonon}. By utilizing a microwave assisted Raman process
involving the upper state $\left\vert f\right\rangle $, the transition of
SiV electronic ground states $\left\vert g\right\rangle $ and $\left\vert
e\right\rangle $ can be effectively coupled to the phononic mode. In this
case, the spin-phonon interaction can be mapped to the Jaynes-Cummings model,
namely
\begin{eqnarray}
H_{s-p} &=&\sum_{n,k}\hbar \omega _{n,k}a_{n,k}^{\dagger }a_{n,k}+\hbar
\omega _{s}\sigma _{ee}  \notag \\
&+&\sum_{n,k}\hbar g_{n,k}(a_{n,k}\sigma _{eg}e^{ikx_{0}}+H.c.),
\end{eqnarray}%
where $\sigma _{ij}=\left\vert i\right\rangle \langle j|$, $\omega _{s}$ is
the effective spin transition frequency, $g\sim 0.1g_{n,k}$, and $g_{n,k}$
is the coupling strength between the SiV center and the phononic modes. Here, we
consider that the defect centers are coupled predominantly to a single band
of the phononic crystal, so the index $n$ be omitted in the following
discussion. In Fig.~$1(d)$, we numerically simulate the corresponding displacement
pattern of the phononic mode by using the finite-element method (FEM), which is performed with the
COMSOL MULTIPHYSICS software.

In a previous work, we proposed the band-gap engineered spin-phonon
interaction. When the spin transition frequency is exactly in a phonon band gap, there will be a phononic bound state. Then we can obtain a much stronger SiV-phononic coupling via tuning
the effective acoustic mode volume \cite{li2019band}. In this context, we now study the
interaction between the phononic crystal modes and an array of SiV spins. Here
we assume that the SiV centers are equally coupled to the phononic mode near
the band gap. Thus the interaction Hamiltonian of the defect spins and  the
phonon modes is expressed as%
\begin{equation}
H_{I}=\sum_{j,k}\hbar g(a_{k}\sigma _{eg}^{j}e^{i\delta _{k}t+ikx_{j}}+H.c.),
\end{equation}%
with $\delta _{k}=\omega _{s}-\omega _{k}$. Assuming the large detuning
regime, $\delta _{k}\gg g$, we can obtain an effective spin-spin interaction
via adiabatically eliminate the phonon modes \cite{james2007effective}. With the
band gap engineered spin-phononic interaction, we integrate over the
phononic modes and obtain the effective Hamiltonian
\begin{equation}
H_{array}=\sum_{i,j}\hbar J_{i,j}\sigma _{eg}^{i}\sigma _{ge}^{j},
\end{equation}%
where
\begin{equation}
J_{i,j}=\frac{g_{c}^{2}}{2\Delta _{BE}}e^{-|x_{i}-x_{j}|/L_{c}}
\end{equation}%
denotes the phononic band-gap mediated spin-spin interaction strength, and $\Delta
_{BE}=\omega _{s}-\omega _{BE}$ is the detuning between the spin transition
and the phononic band edge frequency. $g_{c}=g\sqrt{2\pi a/L_{c}}$ corresponds
to the spin-phononic coupling strength, with $a$ the lattice constant and $%
L_{c}$ the localized length of phononic wavefunction. Going back to the
two-dimensional setup shown in Fig.~$1(a)$, we consider a phononic network
with square lattices on the $x$-$y$ plane, with $2N\times 2N$ SiV spins
located separately at the nodes of the phononic structure. Hence, the
phononic mediated spin-spin interactions can be obtained as%
\begin{eqnarray}
H_{array}^{(2D)} &=&H_{array}^{(x)}+H_{array}^{(y)},  \notag \\
H_{array}^{(x)} &=&\overset{2N}{\sum_{l=1}}\overset{2N}{\sum_{i,j=1}}\hbar
(J_{i,j}\sigma _{eg}^{(i,l)}\sigma _{ge}^{(j,l)}+H.c.),  \notag \\
H_{array}^{(y)} &=&\overset{2N}{\sum_{j=1}}\overset{2N}{\sum_{k,l=1}}\hbar
(J_{k,l}\sigma _{eg}^{(j,k)}\sigma _{ge}^{(j,l)}+H.c.),
\end{eqnarray}%
where $H_{array}^{(x)}$ and $H_{array}^{(y)}$ describe the effective
spin-spin interactions in the $x$ and $y$ directions, respectively. $%
J_{i,j}$ and $J_{k,l}$ are the corresponding phonon mediated spin-spin hopping
rates.

Note that different from the conventional dipole-dipole interaction mediated
by a mechanical resonator or waveguide, this band-gap mediated spin-spin
interaction is decay exponentially with the distance between spins, with a
decay length $L_{c}$. This form of interparticle coupling ($J_{i,j}\sim
e^{-|x_{i}-x_{j}|/\lambda }$) is commonly encountered in several other
quantum systems, such as quantum dot and trapped-ion setups \cite{PhysRevLett.119.210401,PhysRevLett.123.126401}. In the
spin-phononic crystal system, owing to the unique band gap structures of the
phononic crystal, we can get strong and tunable spin-spin interactions by
controlling the mediated phononic modes.

\section{1D topological properties}

\subsection{The periodic driving}

The periodic driving is known to render effective Hamiltonian in which
specific terms can be adiabatically eliminated. In particular, the periodic
driving can be used to trigger nonequilibrium topological behavior in a
trivial setup, which offers an efficient tool to simulate topological
phases in quantum systems \cite{goldman2014periodically,gomez2013floquet}. We consider a periodic
driving quantum system with $H(t)=H(t+T)$, characterized by time period $%
T=2\pi /\omega $. In this case we can introduce Floquet theorem to
investigate long-time dynamics of the system, as developed in Ref. \cite{eckardt2015high}.
With the Floquet-Bloch ansatz, the time-dependent Schrodinger equation be
given by
\begin{equation}
i\hbar d_{t}\left\vert \psi _{\alpha }(t)\right\rangle =H(t)\left\vert \psi
_{\alpha }(t)\right\rangle ,
\end{equation}%
where%
\begin{equation}
\left\vert \psi _{\alpha }(t)\right\rangle =\left\vert \phi _{\alpha
}(t)\right\rangle e^{-i\epsilon _{\alpha }t/\hbar }=e^{-i\epsilon _{\alpha
}t/\hbar }\sum_{m}e^{-im\omega t}\phi _{m}.
\end{equation}%
$\left\vert \psi _{\alpha }(t)\right\rangle $ is the so-called Floquet
eigenstate, and $\epsilon _{\alpha }$ is the quasienergy with band index $%
\alpha $. $\phi _{\alpha }(t)=\phi _{\alpha }(t+T)$ denotes the
time-periodic Floquet eigenmode, which can be constructed by a complete set
of orthonormal basis state $\phi _{m}$. With respect to the basis $%
\left\vert \psi _{\alpha }(t)\right\rangle $, the system can be effectively
described by the Hamiltonian
\begin{equation}
H_{eff}^{mn}=\frac{1}{T}\int_{0}^{T}dte^{i(m-n)\omega t}H(t)\text{.}
\end{equation}%
The effective Hamiltonian is the time-average of the Hamiltonian $H(t)$ in a
driving period, which is the core of Floquet theorem. Note that the Floquet
state in time-periodically driven systems is analogous to the Bloch state in
spatially periodic systems.

In a recent work \cite{PhysRevResearch.2.013121}, we proposed a periodic driving protocol to simulate
topological phases with a color center-phononic crystal system. By applying
a standing wave field between the two lowest sublevels ($|g\rangle
,|e\rangle $) of the SiV center, we get the Floquet engineering of the
spin-spin interactions, resulting in the well-known SSH-type Hamiltonian. Here we
consider a fundamentally different driving protocol, which allows us to selectively control
the spin-spin interactions. What is more important, the resulting spin-spin interactions possess chiral symmetry and
support rich quantum phases associated
with topological invariants.
The time-periodic driving has form \cite{PhysRevLett.123.126401}
\begin{equation}
H_{driv}(t)=\sum_{j}\hbar V_{j}f(t)\sigma _{j}^{z},
\end{equation}%
where $\sigma _{j}^{z}=|e\rangle _{j}\langle e|-|g\rangle _{j}\langle g|$ is
the Pauli operator component. $f(t)$ denotes the standard square-wave
function%
\begin{eqnarray}
f(t) &=&-1\text{ \ \ \ for }t\in \lbrack 0,\frac{T}{2}],  \notag \\
f(t) &=&1\text{ \ \ \ \ for }t\in \lbrack \frac{T}{2},T].
\end{eqnarray}%
$V_{j}$ denotes the on-site potential
\begin{equation}
V_{j}=\left\{
\begin{array}{l}
b_{0}+\frac{(a_{0}+b_{0})}{2}(j-1)\ \ \ \ j=1,3,5,7,... \\
\frac{(a_{0}+b_{0})}{2}j\ \ \ \ \ \ \ \ \ \ \ \ \ \ j=2,4,6,8,...%
\end{array}%
\right. .
\end{equation}%
This stair-like form offers alternating potential difference between\ two
adjacent spins, i.e., $V_{j}-V_{j-1}=a_{0}$ and $V_{j+1}-V_{j}=b_{0}$ are
staggered along the spin array.

We first consider the interaction of the periodic driving and
the 1D spin array, but the case of 2D will be studied in the next
section. Now we transform the total Hamiltonian $%
H_{1D}=H_{array}+H_{driv}(t) $ into the interaction picture, with the
unitary operator $U(t)=e^{-i\int_{0}^{t}d\tau H_{driv}(\tau )/\hbar }$.
After the unitary transformation, we obtain
\begin{eqnarray}
\sigma _{eg}^{j} &\rightarrow &e^{i\Delta _{j}(t)\sigma _{j}^{z}}\sigma
_{eg}^{j}e^{-i\Delta _{j}(t)\sigma _{j}^{z}}=\sigma _{eg}^{j}e^{2i\Delta
_{j}(t)},  \notag \\
\sigma _{ge}^{j} &\rightarrow &e^{i\Delta _{j}(t)\sigma _{j}^{z}}\sigma
_{ge}^{j}e^{-i\Delta _{j}(t)\sigma _{j}^{z}}=\sigma _{ge}^{j}e^{-2i\Delta
_{j}(t)},
\end{eqnarray}%
with%
\begin{eqnarray}
\Delta _{j}(t) &=&V_{j}\int_{0}^{t}d\tau f(\tau )  \notag \\
&=&V_{j}\int_{0}^{t}d\tau \lbrack \underset{n\neq 0}{\sum }\frac{1}{n\pi i}%
(e^{-in\pi }-1)e^{in\omega \tau }],
\end{eqnarray}%
where we expanded $f(t)$ into its Fourier series. In the interaction
picture, the total Hamiltonian has the form as
\begin{equation}
H_{1D}=\underset{i,j}{\sum }\hbar J_{ij}(t)\sigma _{eg}^{i}\sigma
_{ge}^{j},
\end{equation}%
where $J_{ij}(t)=J_{ij}e^{2i(\Delta _{i}(t)-\Delta _{j}(t))}$ is the hopping
rate with a temporal periodicity, $J_{ij}(t)=J_{ij}(t+T)$. The Floquet
components of the Hamiltonian ($14$) read%
\begin{equation}
H_{1D}^{mn}=\underset{i,j}{\sum }\hbar J_{ij}^{mn}\sigma _{eg}^{i}\sigma
_{ge}^{j},
\end{equation}%
\begin{equation}
J_{ij}^{mn}=\frac{1}{T}\int_{0}^{T}dtJ_{ij}(t)e^{i(m-n)\omega t}.
\end{equation}%
For the time-periodically driven system, $H_{1D}^{mn}$ can be expressed by
the Floquet-Magnus expansion. In the high-frequency regime $\omega \gg
J_{i,j}$, it is a good approximation to neglect the rapid oscillation of the
external driving \cite{rahav2003effective,eckardt2015high,mikami2016brillouin}.  As a result, the spin-spin interaction can be given by the
zeroth-order expansion term
\begin{equation}
\mathcal{J}_{ij}=J_{ij}\frac{i\omega }{2\pi (V_{i}-V_{j})}(e^{-i2\pi
(V_{i}-V_{j})/\omega }-1).
\end{equation}

\begin{figure}[tbp]
\includegraphics[width=8.6cm]{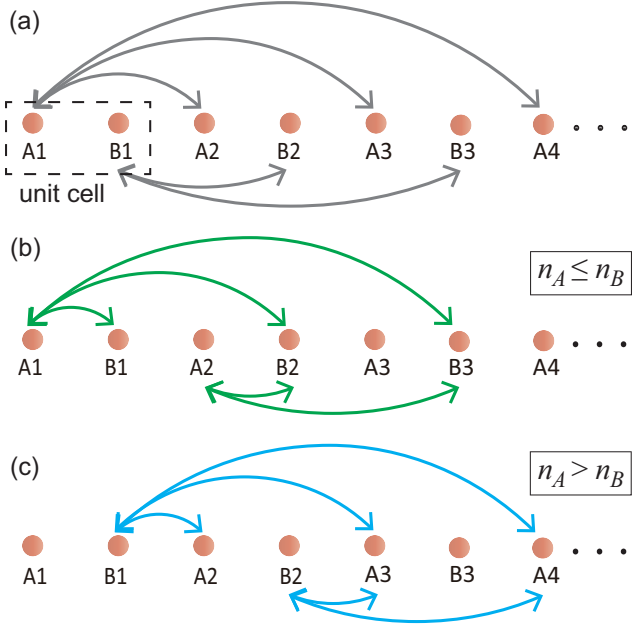}
\caption{(Color online) Schematic diagram for effective spin-spin
interactions. (a) Even-neighbor hopping. (b) and (c) illustrate two kinds of
odd-neighbor hopping examples, respectively.}
\label{fig_effective}
\end{figure}

As for the SSH model, the interparticle interaction is characterized by
staggering hopping amplitudes. Thus, the two nearest-neighbor spins can be
grouped into a unit cell and classified as odd and even spins, as we
proposed in Ref. \cite{PhysRevResearch.2.013121}. Likewise, here we consider the phononic-mediated
spin-spin interaction as a bipartite lattice of the form $ABABAB$, with
\begin{eqnarray}
A_{n} &=&\sigma _{ge}^{j}\ \ \ j=1,3,5,7,...,  \notag \\
B_{n} &=&\sigma _{ge}^{j}\ \ \ j=2,4,6,8,....
\end{eqnarray}%
Based on this definition, we rewrite the renormalized hopping amplitude $%
\mathcal{J}_{ij}$. For simplicity, here we introduce $n_{l}$ to label the spins
at the $l$ site of the $n$th cell, $l=A$ or $B$. In general, there are two
types of interparticle hopping. For the even-neighbor hopping, which
describes the spin-spin interaction of the same sublattice, as shown in Fig.
$2(a)$. The potential difference are $V_{i}-V_{j}=\pm m(a_{0}+b_{0})$, with $%
m=n_{l}^{\prime }-n_{l}$. In consequence, the even-neighbor spin-spin
hopping rate can be written as
\begin{equation}
\mathcal{J}_{n_{l},n_{l}^{\prime }}=\frac{iJ_{n_{l},n_{l}^{\prime }}}{\mp
2\pi qm}(e^{\pm i2\pi qm}-1),
\end{equation}%
where $q=(a_{0}+b_{0})/\omega $, and \textquotedblleft $\pm $" correspond to$\ $%
the coupling to the right and left spins, respectively. From Eq. (20), the
even-neighbor hopping is always zero if we assign $q=1,2,3,...$. Thus we
conclude that the even-neighbor hopping can be suppressed by tuning the
parameters $a_{0}$ and $b_{0}$. Note that the even-neighbor hopping is a
detrimental source for the chiral symmetry \cite{PhysRevLett.123.126401}.

For the odd-neighbor hopping, which describes the spin-spin interaction of
the different sublattice. To better describe the physical picture of the
spin-spin interaction, we further classify two kinds of odd-neighbor
hopping. We first discuss the case with $n_{A}\leq n_{B}$, for which  the schematic
diagram is shown in Fig. $2(b)$. If we define $n_{B}=n_{A}+r$ $(r=0,1,2,...)$%
, the spin-spin interaction can be described by
\begin{eqnarray}
\mathcal{J}_{n_{A},n_{B}} &=&-\frac{iJ_{n_{A},n_{B}}}{2\pi (qr+\frac{a_{0}}{%
\omega })}[e^{2i\pi (qr+\frac{a_{0}}{\omega })}-1],  \notag \\
\mathcal{J}_{n_{B},n_{A}} &=&\frac{iJ_{n_{A},n_{B}}}{2\pi (qr+\frac{a_{0}}{%
\omega })}[e^{-2i\pi (qr+\frac{a_{0}}{\omega })}-1],
\end{eqnarray}%
where $\mathcal{J}_{n_{A},n_{B}}$ and $\mathcal{J}_{n_{B},n_{A}}$ describe
the forward $(A\rightarrow B)$ and backward $(B\rightarrow A)$ hoppings,
respectively. For the case with $n_{A}>n_{B}$, the schematic diagram is
shown in Fig. $2(c)$. If we define $n_{B}=n_{A}-r^{\prime }$ $(r^{\prime
}=1,2,3,...)$, the spin-spin interaction can be described by
\begin{eqnarray}
\mathcal{J}_{n_{A},n_{B}}^{\prime } &=&\frac{iJ_{n_{A},n_{B}}}{2\pi
(qr^{\prime }-\frac{a_{0}}{\omega })}[e^{-2i\pi (qr^{\prime }-\frac{a_{0}}{%
\omega })}-1],  \notag \\
\mathcal{J}_{n_{B},n_{A}}^{\prime } &=&-\frac{iJ_{n_{A},n_{B}}}{2\pi
(qr^{\prime }-\frac{a_{0}}{\omega })}[e^{2i\pi (qr^{\prime }-\frac{a_{0}}{%
\omega })}-1].
\end{eqnarray}%
Likewise, $\mathcal{J}_{n_{A},n_{B}}^{\prime }$ and $\mathcal{J}%
_{n_{B},n_{A}}^{\prime }$ represent the backward $(A\rightarrow B)$ and
forward $(B\rightarrow A)$ hopping, respectively. From Eqs. (20) and (21),
we can conclude
\begin{equation}
\mathcal{J}_{n_{A},n_{B}}=(\mathcal{J}_{n_{B},n_{A}})^{\ast },\mathcal{J}%
_{n_{A},n_{B}}^{\prime }=(\mathcal{J}_{n_{B},n_{A}}^{\prime })^{\ast }.
\end{equation}%
Unlike the case of the SSH model, the backward and forward hoppings of the odd-neighbor
spin-spin interaction are not equal.

According to this bipartite solution, the Hamiltonian $H_{1D}$ can be
rewritten as
\begin{eqnarray}
H_{1D} &=&\underset{n,r,r^{\prime }}{\sum }\hbar (\mathcal{J}%
_{n_{A},n_{B}}A_{n}B_{n+r}^{\dagger }+\mathcal{J}_{n_{B},n_{A}}A_{n}^{%
\dagger }B_{n+r}  \notag \\
&&+\mathcal{J}_{n_{A},n_{B}}^{\prime }A_{n}B_{n-r^{\prime }}^{\dagger }+%
\mathcal{J}_{n_{B},n_{A}}^{\prime }A_{n}^{\dagger }B_{n-r^{\prime }}).
\end{eqnarray}%
Here we neglected the even-neighbor hopping terms. By applying a particular
periodic driving field to the SiV centers, we obtain the Floquet engineering of
the spin-spin interactions with unique properties. In this case, the even-neighbor hopping is
suppressed by tuning the parameters of the driving field, while the odd-neighbor
hopping can be enhanced as needed. This scheme enforces the chiral symmetry
which provides topological protection for the spin-spin interaction.
In Fig. $%
3$, we numerically calculate the quasienergy spectrum as a function of $%
a_{0} $. We can see that all the eigenmodes are grouped into chiral symmetric
pairs with opposite energies. For simplicity, here we express the bare
spin-spin interaction as
\begin{equation}
J_{i,j}=\frac{g_{c}^{2}}{2\Delta _{BE}}%
e^{-|x_{i}-x_{j}|/L_{c}}=J_{0}e^{-|x_{i}-x_{j}|/L_{c}}.
\end{equation}%
Given that the band-gap mediated spin-spin interaction decays
exponentially with the spin spacing, here only the first- and third-
neighbor interactions are included.
\begin{figure}[tbp]
\includegraphics[width=8.6cm]{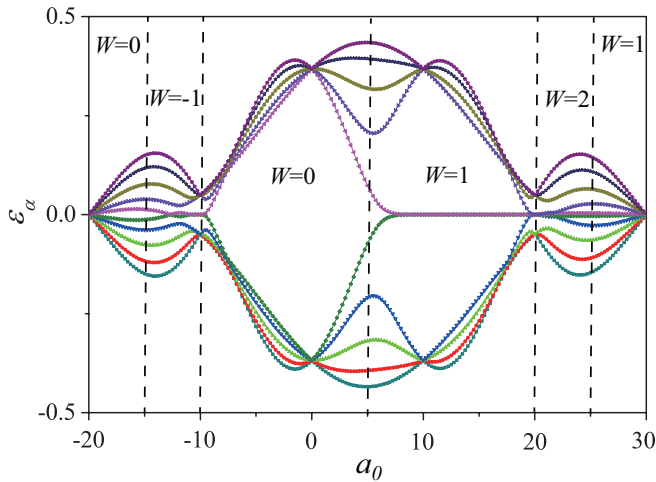}
\caption{(Color online) (a) Quasienergy spectrum as a function of $a_{0}$,
the corresponding winding number $\mathcal{W}$ is indicated. Here we assign $%
|x_{j}-x_{j+1}|=a$, $L_{c}=a$. Note that only the first and third
odd-neighbor interactions are included. The other parameters are $N=5$, $%
\protect\omega =10$, $q=1$ and $J_{0}=1$.}
\label{fig_energy}
\end{figure}

\subsection{Topological phases}
\begin{figure*}[tbp]
\includegraphics[width=18cm]{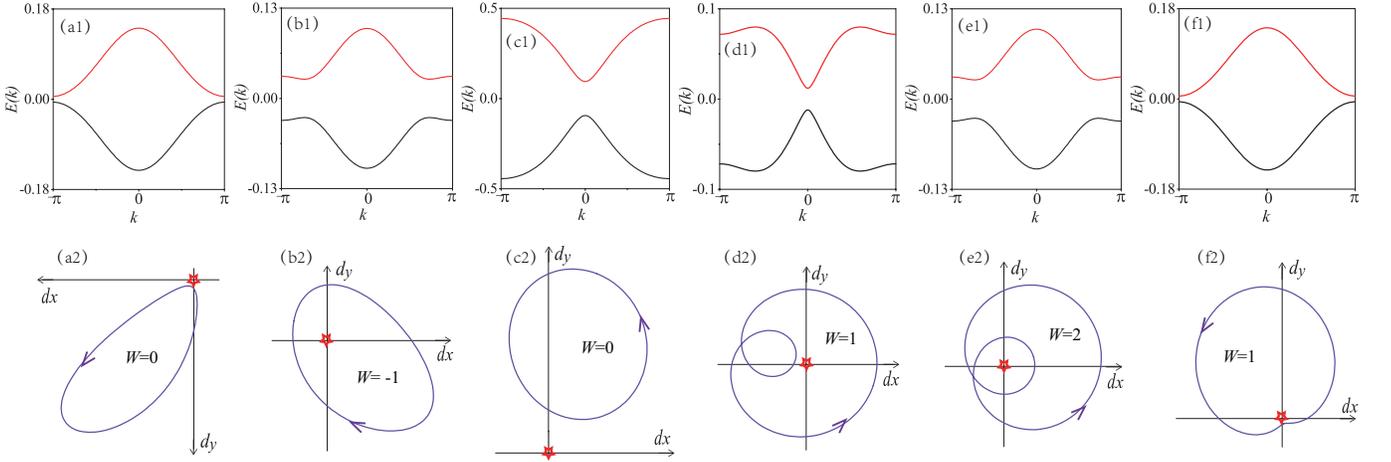}
\caption{(Color online) (a1)-(f1) show the dispersion relations for various
parameter settings of periodic driving: (a1) $a_{0}=-16$. (b1) $a_{0}=-11$.
(c1) $a_{0}=4$. (d1) $a_{0}=19$. (e1) $a_{0}=21$. (f1) $a_{0}=26$. As the
wave number $k$ runs through the Brillouin zone ($-\protect\pi $, $\protect%
\pi $), the energy spectrum splits into two branches and there exists a band
gap between the lower and higher branches. (a2)-(f2) correspond to the
winding configuration of $(d_{x},d_{y})$ around the origin (red star),
and the relevant winding number $\mathcal{W}$ is explicitly shown.  In (a2) and (c2),
the loop wind avoids the origin, and then $\mathcal{W}=0$. In (b2), (d2) and
(f2), the endpoint of $d(k)$ encircles the origin once, but these are
topological inequivalent. For (d2) and (f2), the endpoint of $d(k)$ is a
closed loop in the counter clockwise direction, and $\mathcal{W}=1$. While in (b2), the
endpoint of $d(k)$ is along the clockwise direction, and $\mathcal{W}=-1$. In (e2), the
endpoint of $d(k)$ encircles the origin two times, and $\mathcal{W}=2$. Other
parameters are the same as those in Fig. $3$. }
\label{fig_energy}
\end{figure*}
The periodic driving protocol offers an
effective method to investigate the topological character of the
spin-phononic crystal system. To explore topological features of the
Floquet engineering spin-spin system, we convert $H_{1D}$ to the momentum space.
Considering periodic boundary conditions, we can make the Fourier
transformation
\begin{equation}
O_{n}=\frac{1}{\sqrt{N}}\sum_{k}e^{ink}O_{k},(O=A,B)
\end{equation}%
where $k=2\pi m/N(m=1,2,...,N)$ is the wavenumber in the first Brillouin
zone, and $A_{k}$ and $B_{k}$ are the momentum space operators. Defining the
unitary operator $\psi(k)=\left(
\begin{array}{cc}
A_{k} & B_{k}%
\end{array}%
\right) ^{T}$, the Hamiltonian $H_{1D}$ be expressed as%
\begin{equation}
H_{1D}=\sum_{k}\psi(k)^{\dagger }H(k)\psi(k).
\end{equation}%
Then we obtain $2\times 2$ matrix form of the Hamiltonian in the $k$-space
\begin{equation}
H(k)=\hbar \left(
\begin{array}{cc}
0 & f(k) \\
f^{\ast }(k) & 0%
\end{array}%
\right) .
\end{equation}%
with%
\begin{equation}
f(k)=\underset{r,r^{\prime }}{\sum }(\mathcal{J}_{n_{B},n_{A}}e^{ikr}+%
\mathcal{J}_{n_{B},n_{A}}^{^{\prime }}e^{-ikr^{\prime }}).
\end{equation}%
Here $f(k)$ describes the coupling between the $A$ and $B$ spins in momentum
space.

The dispersion relation can be obtained by solving the eigenvalue
equation
\begin{equation}
H(k)\psi (k)=E(k)\psi (k),
\end{equation}%
using the fact that $H^{2}(k)=E^{2}(k)I$, with $I$ being the identity
operator in the Hilbert space. Then we obtain the energy band structure as
\begin{equation}
E(k)=\pm \hbar |f(k)|,
\end{equation}%
\begin{equation}
\psi (k)=\frac{1}{\sqrt{2}}\binom{1}{\pm e^{-i\vartheta (k)}}.
\end{equation}
$\psi (k)$ corresponds to the eigenfunctions for the lower and upper band,  and $%
\vartheta (k)$ is defined as the argument of $f(k)$. Figs. 4(a1)-(f1) show
the energy spectra for different driving field parameters, which are split
into two branches and there exists a band gap between the lower and upper
branches. It should be noticed that the band gap will be vanished at the
critical point of topological phases.

The band gap structures are generally associated with topological properties
of bulk-boundary correspondence. For the 1D Floquet engineering spin-spin system, we
introduce the topological Zak phase \cite{shen2018topological}
\begin{eqnarray}
\mathcal{\varphi }_{Zak} &=&-i\overset{occ.}{\underset{j=1}{\sum }}%
\int_{0}^{2\pi }dk\psi ^{\dagger }(k)\partial _{k}\psi (k) \notag \\
&=&N_{occ.}\frac{1}{2}\int_{0}^{2\pi }dk\frac{d}{dk}\vartheta (k) \notag \\
&=&\mathcal{W}\pi
\end{eqnarray}%
where $\mathcal{W}$ is the topological winding number, $N_{occ.}$ describes
the number of occupied energy bands. Now we need to investigate the winding
number of the system. Alternatively, $f(k)$ can be expressed in the form
\begin{equation}
f(k)=d(k)\cdot \sigma ,
\end{equation}%
where $\sigma =(\sigma _{x},\sigma _{y},\sigma _{z})$ is the Pauli matrix,
and $d(k)$ denotes a three-dimensional vector field
\begin{eqnarray}
d_{x}(k) &=&\frac{1}{2}\underset{r,r^{\prime }}{\sum }(\mathcal{J}_{n_{B},n_{A}}e^{ikr}+%
\mathcal{J}_{n_{B},n_{A}}^{^{\prime }}e^{-ikr^{\prime }}+c.c.),  \notag \\
d_{y}(k) &=&\frac{i}{2}\underset{r,r^{\prime }}{\sum }(\mathcal{J}_{n_{B},n_{A}}e^{ikr}+%
\mathcal{J}_{n_{B},n_{A}}^{^{\prime }}e^{-ikr^{\prime }}-c.c.),  \notag \\
d_{z}(k) &=&0.
\end{eqnarray}%

For general $2$-band topological insulators, owing to the periodicity of the
momentum-space Hamiltonian, the path of the endpoint of $d(k)$ is a closed
loop in the auxiliary space $(d_{x},d_{y})$ \cite{asboth2016short}. The topology of this loop can
be characterized by an integer, the winding number%
\begin{equation}
\mathcal{W}=\frac{1}{2\pi }\int_{0}^{2\pi }\mathbf{n}\times \partial _{k}\mathbf{n}dk,
\end{equation}%
where $\mathbf{n}=(n_{x},n_{y})=(d_{x},d_{y})/\sqrt{d_{x}^{2}+d_{y}^{2}}$ is the
normalized vector. Here the winding number $\mathcal{W}$ counts the number
of times the loop winds around the origin of the $dx$-$dy$ plane. Figs.
4(a2)-(f2) present the path of the endpoint of $d(k)$ on the $dx$-$dy$
plane. For different values of $a_{0}$, the winding number of the system
exhibits four possible values, $-1$, $0$, $1$, $2$. According to Eq. (33),
we can derive the relevant topological Zak phases directly. Furthermore, one
can implement the topological phase transition in this SiV-phononic system
by modulating the periodic driving.

From the numerical simulation results, we show rich quantum  phases related to topological invariants. As for the generalized SSH model, a prototypical example to investigate topological
properties in a trivial system, the 1D Zak phase has only two possible values $%
0$ or $\pi $. This work offers an effective scheme for studying topological phases induced by periodic driving. The distinct feature is that it enables to simulate higher-order topological phases and related topological phase transitions in
topological trivial systems.

\subsection{Edge states}
\begin{figure*}[tbp]
\includegraphics[width=18cm]{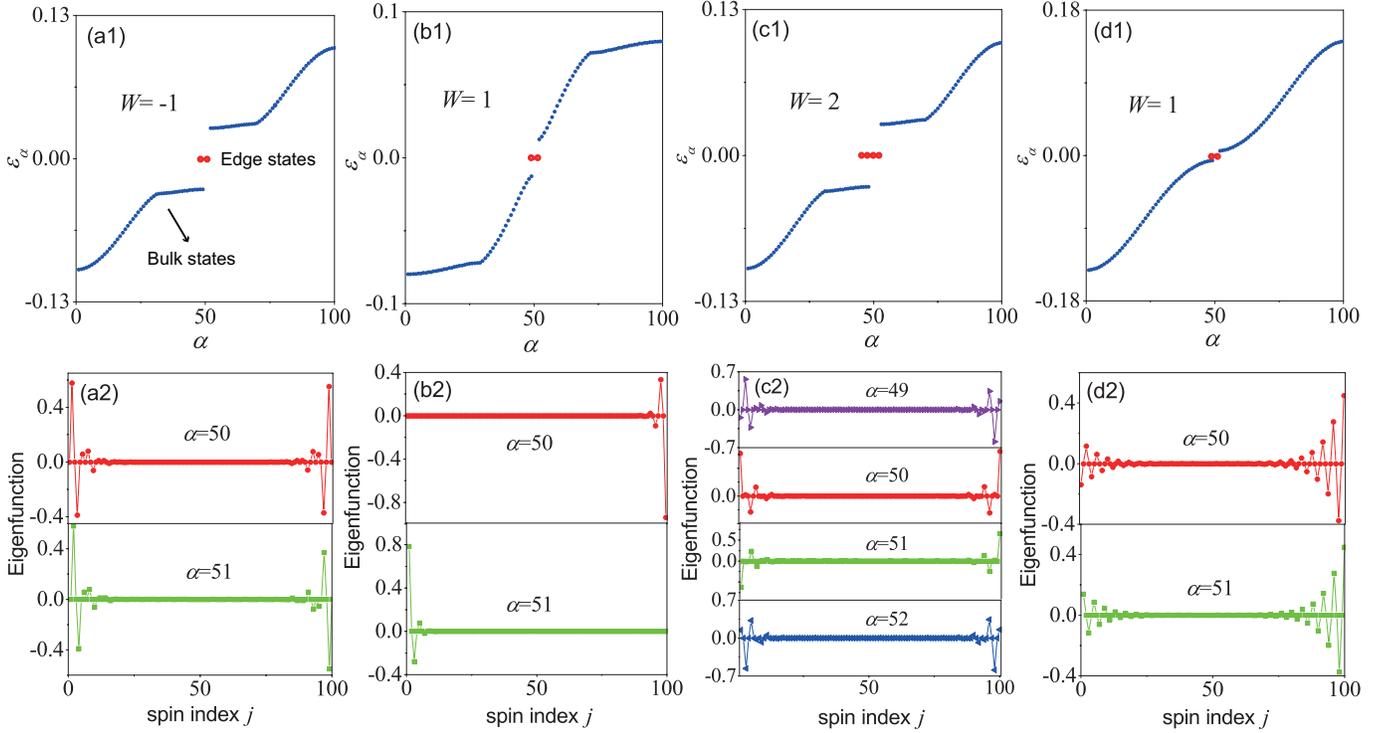}
\caption{(Color online) (a1)-(d1) show the energy spectrum in the
single-excited state subspace for various parameter settings of periodic
driving: (a1) $a_{0}=-11$. (b1) $a_{0}=19$. (c1) $a_{0}=21$. (d1) $a_{0}=26$%
. For the cases with $\mathcal{W}=1$ and $\mathcal{W}=-1$, there are two
zero-energy eigenvalues. For the case with $\mathcal{W}=2$, there are four
zero-energy eigenvalues. The zero-energy eigenvalues (the red point)
correspond to the topological edge states. The rest of eigenvalues (the blue
point) correspond to the bulk states of the system. (a2)-(d2) show the
related eigenfunction of the gapless modes. Here we consider $N=50$. Other
parameters are the same as those in Fig. $3$. }
\label{fig_energy}
\end{figure*}
The existence of edge states at the boundary is a distinguished feature for
topological insulator states. In the following, we first
simulate the edge states in a 1D spin-phononic system. The core
step is to look for the zero-energy eigenstates. Here we introduce the
single-excited state
\begin{equation}
\psi =\underset{n}{\sum }(a_{n}A^{\dagger }_{n}+b_{n}B^{\dagger
}_{n})\left\vert 0\right\rangle,
\end{equation}%
where $a_{n}$ and $b_{n}$ are the amplitudes of occupying probability in the
$n$th cell. $\left\vert 0\right\rangle=\left\vert ggg...\right\rangle$ is the vacuum state, which describes that all spins stay in the ground state $%
|g\rangle $. In the single-excited state subspace, we can get the the
zero-energy eigenstates by sloving
\begin{equation}
H_{1D}\underset{n}{\sum }(a_{n}A^{\dagger }_{n}+b_{n}B^{\dagger
}_{n})\left\vert 0\right\rangle=0.
\end{equation}%
There will be $2N$ equations for the amplitudes $a_{n}$ and $%
b_{n}$. Considering the boundary conditions, $b_{0}=a_{N+1}=0$. We can analytically derive the left and right zero-energy edge
states, respectively.

To verify the model, we numerically simulate the energy spectrum and
zero-energy eigenstates of the system. Figs.~5(a1)-(d1) show the eigenvalues
for various parameter settings of the periodic driving. As for the
non-topological regime, $\mathcal{W}=0$, there will be an energy band gap
but no gapless modes appear. For the cases with $\mathcal{W}=1$ and $%
\mathcal{W}=-1$, there are two zero-energy eigenvalues. For the case with $%
\mathcal{W}=2$, there are four zero-energy eigenvalues. Correspondingly, we
plot the zero-energy edge states in Figs.~5(a2)-(d2). We see that the
wavefunctions are located at the vicinity of the array boundaries, which are
the so-called topological edge states. In addition, the edge states only
distribute at certain (odd or even) sites, which is related to the chiral
symmetry of the system.

\section{2D topological properties}

\subsection{The periodic driving}

Now we proceed to generalize the above 1D results to 2D spin-phononic
crystal networks. Here we consider adding two mutually perpendicular
microwave fields to the color center arrays \cite{zou2017quantum}. The first
one is a time-dependent microwave driving of frequency $\omega _{x}$ in the $%
x$ direction. The other is a time-dependent driving of frequency $\omega
_{y} $ in the $y$ direction. These two periodic driving fields have the form
\begin{eqnarray}
H_{driv}^{(x)} &=&\overset{2N}{\sum_{l=1}}\overset{2N}{\sum_{j=1}}\hbar
V_{j,l}f_{x}(t)\sigma _{j,l}^{z},  \notag \\
H_{driv}^{(y)} &=&\overset{2N}{\sum_{j=1}}\overset{2N}{\sum_{l=1}}\hbar
V_{j,l}f_{y}(t)\sigma _{j,l}^{z}.
\end{eqnarray}%
$V_{j,l}=(V_{j},V_{l})$ describes the on-site potential in the 2D phononic
network, the two components of which are in the form of Eq.~$(12)$. $f_{x}(t)$ and $%
f_{y}(t)$ denote the square-wave function in the $x$ and $y$ directions,
respectively.

\begin{figure}[tbp]
\includegraphics[width=8.6cm]{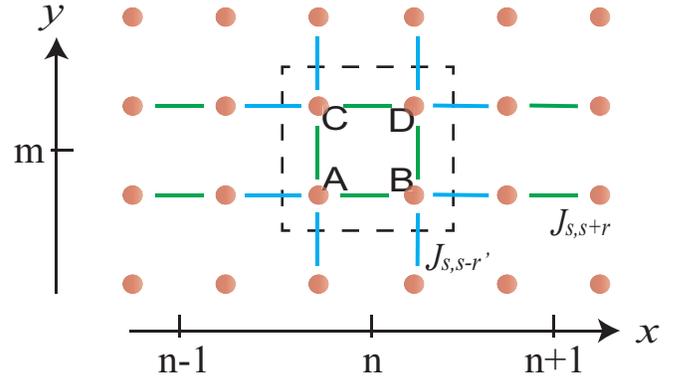}
\caption{(Color online) Schematic diagram of the 2D Floquet engineering spin-spin interaction. There are four spins in a unit cell, which are labeled as \{$A,B,C,D$\}, respectively. Here we introduce $(n,m)$ to describe the position of each
unit cell in the 2D spin-spin networks. For simplicity, only the nearest-neighbor interactions are illustrated.}
\label{fig_energy}
\end{figure}
Let us discuss the two directions separately. For the periodic driving spin
arrays along the $x$ direction, the total Hamiltonian can be written as%
\begin{equation}
H_{2D}^{(x)}=H_{array}^{(x)}+H_{driv}^{(x)}.
\end{equation}%
In the interaction picture, we introduce the unitary operator $%
U_{x}(t)=e^{-i\int_{0}^{t}d\tau H_{driv}^{(x)}/\hbar }$. After the unitary
transformation, we obtain%
\begin{equation}
\sigma _{eg}^{(j,l)}\rightarrow \sigma _{eg}^{(j,l)}e^{2i\Delta
_{j}(t)},\sigma _{ge}^{(j,l)}\rightarrow \sigma _{ge}^{(j,l)}e^{-2i\Delta
_{j}(t)},
\end{equation}%
with%
\begin{equation}
\Delta _{j}(t)=V_{j}\int_{0}^{t}d\tau f(\tau ).
\end{equation}%
While for the spin arrays along the $y$ direction, the total Hamiltonian is
given by%
\begin{equation}
H_{2D}^{(y)}=H_{array}^{(y)}+H_{driv}^{(y)}.
\end{equation}%
Similary, we introduce the unitary operator $U_{y}(t)=e^{-i\int_{0}^{t}d\tau
H_{driv}^{(y)}/\hbar }$. After the unitary transformation, we obtain%
\begin{equation}
\sigma _{eg}^{(j,l)}\rightarrow \sigma _{eg}^{(j,l)}e^{2i\Delta
_{l}(t)},\sigma _{ge}^{(j,l)}\rightarrow \sigma _{ge}^{(j,l)}e^{-2i\Delta
_{l}(t)},
\end{equation}%
with%
\begin{equation}
\Delta _{l}(t)=V_{l}\int_{0}^{t}d\tau f(\tau ).
\end{equation}%
In the following, analogous to the 1D case, we consider the
bipartite interaction in both $x$ and $y$ directions. Then we get a
2D system with $N\times N$ unit cells. As depicted in Fig.~$6$,
there are four spins in each unit cell, which are labeled as \{$A,B,C,D$\},
respectively. In the regime $\omega \gg J_{i,j},J_{k,l}$, we derive the
Floquet engineering spin-spin interactions along the $x$ and $y$ directions%
\begin{eqnarray}
H_{2D}^{(x)} &=&\underset{m}{\sum }\underset{n,r,r^{\prime }}{\sum }\hbar
\lbrack \mathcal{J}_{n,n+r}(A_{n,m}B_{n+r,m}^{\dagger
}+C_{n,m}D_{n+r,m}^{\dagger })  \notag \\
&&+\mathcal{J}_{n,n-r^{\prime }}^{\prime }(A_{n,m}B_{n-r^{\prime
},m}^{\dagger }+C_{n,m}D_{n-r^{\prime },m}^{\dagger })+H.c.],  \notag \\
H_{2D}^{(y)} &=&\underset{n}{\sum }\underset{m,r,r^{\prime }}{\sum }\hbar
\lbrack \mathcal{J}_{m,m+r}(A_{n,m}C_{n,m+r}^{\dagger
}+B_{n,m}D_{n,m+r}^{\dagger })  \notag \\
&&+\mathcal{J}_{m,m-r^{\prime }}^{\prime }(A_{n,m}C_{n,m-r^{\prime
}}^{\dagger }+B_{n,m}D_{n,m-r^{\prime }}^{\dagger })+H.c.].
\end{eqnarray}%
For simplicity, we introduce  $(n,m)$ to describe the position of each
unit cell in the 2D spin-spin networks, with $n,m=1,2,\ldots ,N$.

To simplify the model, here we suppose that the spin spacing $d_{x}=d_{y}$
and the periodic driving frequencies $\omega _{x}=\omega _{y}$. In this case, we
can derive
\begin{eqnarray}
\mathcal{J}_{n,n+r} &=&\mathcal{J}_{m,m+r},  \notag \\
\mathcal{J}_{n,n-r^{\prime }}^{\prime } &=&\mathcal{J}_{m,m-r^{\prime
}}^{\prime }.
\end{eqnarray}%
Thus we can define $s=n$ or $m$, and the two-dimensional Hamiltonian can be further
integrated as%
\begin{eqnarray}
H_{2D} &=&\underset{r,r^{\prime }}{\sum }\underset{n,m}{\sum }\hbar \lbrack
\mathcal{J}_{s,s+r}(A_{n,m}B_{n+r,m}^{\dagger }+C_{n,m}D_{n+r,m}^{\dagger }
\notag \\
&&+A_{n,m}C_{n,m+r}^{\dagger }+B_{n,m}D_{n,m+r}^{\dagger })  \notag \\
&&+\mathcal{J}_{s,s-r^{\prime }}^{\prime }(A_{n,m}B_{n-r^{\prime
},m}^{\dagger }+C_{n,m}D_{n-r^{\prime },m}^{\dagger }  \notag \\
&&+A_{n,m}C_{n,m-r^{\prime }}^{\dagger }+B_{n,m}D_{n,m-r^{\prime }}^{\dagger
})+H.c.].
\end{eqnarray}

\subsection{Topological phases}

To investigate the topological features in the 2D Floquet engineering spin-spin
system, we convert the Hamiltonian $H_{2D}$ to the momentum space. Here we
consider periodic boundary conditions along both the $x$ and $y$ directions.
Then we apply the Fourier transformation to the four spins in a unit cell
\begin{equation}
O_{n,m}=\frac{1}{\sqrt{N}}\sum_{\mathbf{k}}e^{i(k_{x}n+k_{y}m)}O_{\mathbf{k}%
},(O=A,B,C,D)
\end{equation}%
where $\mathbf{k}=(k_{x},k_{y})$ is the wavenumber in the first Brillouin
zone. If we define the unitary operator $\psi (\mathbf{k})=\left(
\begin{array}{cccc}
A_{\mathbf{k}} & B_{\mathbf{k}} & C_{\mathbf{k}} & D_{\mathbf{k}}%
\end{array}%
\right) ^{T}$, the two-dimensional Hamiltonian can be rewritten as%
\begin{equation}
H_{2D}=\sum_{\mathbf{k}}\psi ^{\dagger }(\mathbf{k})H(\mathbf{k})\psi (%
\mathbf{k}).
\end{equation}%
Along with that we get $4\times 4$ matrix form of the Hamiltonian in the $%
\mathbf{k}$-space
\begin{equation}
H(\mathbf{k})=\hbar\left(
\begin{array}{cccc}
0 & f(k_{x}) & f(k_{y}) & 0 \\
f^{\ast }(k_{x}) & 0 & 0 & f(k_{y}) \\
f^{\ast }(k_{y}) & 0 & 0 & f(k_{x}) \\
0 & f^{\ast }(k_{y}) & f^{\ast }(k_{x}) & 0%
\end{array}%
\right) .
\end{equation}%
$f(k_{x})$ and $f(k_{y})$ have the same form as Eq. (29),\ which describe
the spin-spin couplings in the $x$ and $y$ directions, respectively.

Let us study the 2D dispersion relation by solving the eigenvalue equation%
\begin{equation}
H(\mathbf{k})\psi (\mathbf{k})=E(\mathbf{k})\psi (\mathbf{k}),
\end{equation}%
then obtain%
\begin{equation}
E(\mathbf{k})=\epsilon _{x}\hbar \left\vert f(k_{x})\right\vert +\epsilon
_{y}\hbar \left\vert f(k_{y})\right\vert ,
\end{equation}%
\begin{equation}
\psi (\mathbf{k})=\frac{1}{2}\left(
\begin{array}{c}
1 \\
\epsilon _{x}e^{-i\vartheta _{x}(k_{x})} \\
\epsilon _{y}e^{-i\vartheta _{y}(k_{y})} \\
\epsilon _{x}\epsilon _{y}e^{-i[\vartheta _{x}(k_{x})+\vartheta _{y}(k_{y})]}%
\end{array}%
\right) ,
\end{equation}%
where $\epsilon _{i}=\pm 1$, $\vartheta _{i}(k_{i})=\arg [f(k_{i})],$ $i=x,y$%
. In Figure.~7(a), we numerically calculate the 2D energy spectrum in the
momentum space. There are four energy bands since there are four spins in a
unit cell. The lowest and highest bands are isolated, while the two middle
bands are jointed at the edges of the Brillouin zone ($0,0$), ($\pm \pi ,\pm
\pi $), ($\mp \pi ,\pm \pi $). According to Eq. (53), there exist two equal
energy band gaps. When assigning suitable values of $a_{0}$, these four bands
will be jointed together, and the band gaps vanished. This is a signature of
topological phase transition.

\begin{figure*}[tbp]
\includegraphics[width=18cm]{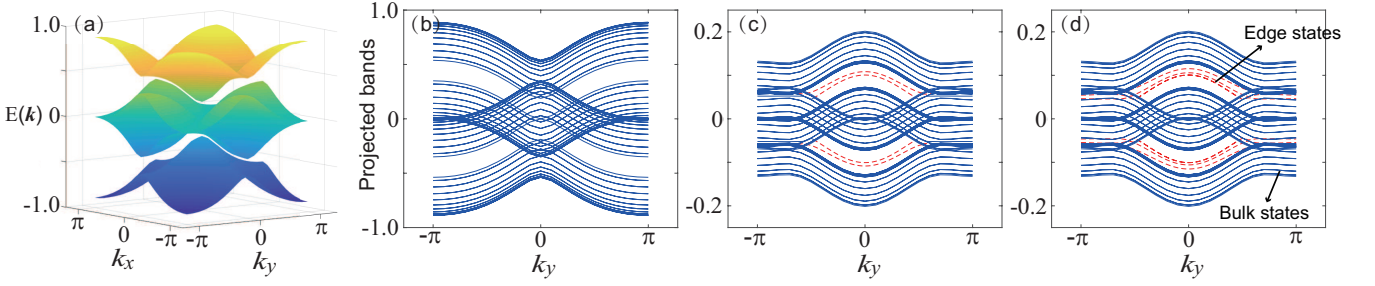}
\caption{(Color online)  (a) Band structure of the 2D Floquet engineering spin-spin interaction in the $\mathbf{k}$-space, $a_{0}=4$. The energy spectrum has four branches. The lowest and highest bands are isolated, while the two middle
bands are touched at the edges of the Brillouin zone ($0,0$), ($\pm \pi ,\pm
\pi $), ($\mp \pi ,\pm \pi $). (b)-(d) show the
projected band structures for various parameter settings of periodic
driving: (b) $a_{0}=4$. (c) $a_{0}=-11$. (d) $a_{0}=21$. The blue and red curves denote the bulk and edge modes, respectively. Here we consider $N_{x}=11$. Other
parameters are the same as those in Fig. $3$. }
\label{fig_energy}
\end{figure*}

For 2D systems, the topological invariants of energy bands are
generally characterized by the Chern number. If we define the Bloch function $%
\psi _{m}(\mathbf{k})$ for the $m$th energy band, the non-Abelian Berry
connection $A_{m}(\mathbf{k})=i\psi _{m}^{\dagger }(\mathbf{k})\partial _{%
\mathbf{k}}\psi _{m}(\mathbf{k})$. The topological Chern number can be calculated
by the integral of $A_{m}(\mathbf{k})$ over the first Brillouin zone,
\begin{equation}
\mathcal{C}=\frac{1}{2\pi }\int_{BZ}d^{2}\mathbf{kTr}[A_{m}(\mathbf{k})].
\end{equation}%
The integral runs over all occupied bands. Alternatively, the Chern number
can be defined by the vector field $d(\mathbf{k})$
\begin{equation}
\mathcal{C}=\frac{1}{4\pi }\int \int dk_{x}dk_{y}(\partial _{k_{x}}\mathbf{n}\times
\partial _{k_{y}}\mathbf{n})\cdot \mathbf{n},
\end{equation}
where $\mathbf{n}=d(\mathbf{k})/\left\vert d(\mathbf{k})\right\vert $. This implies that the topological invariant Chern number can be determined from the
winding number in momentum space. Therefore, for this 2D periodically driving
spin-spin interactions, the Chern number has four values, $-\frac{1}{2},0,%
\frac{1}{2},1$. It should be noted that the Chern number here is not quantized as an integer multiple, which is different from the traditional concept. For this reason, some works introduce a polarization vector to describe the topological invariant in the 2D system \cite{PhysRevA.100.032102,PhysRevB.100.075120,PhysRevLett.118.076803,PhysRevLett.122.233903}. As mentioned above, we consider the square lattice geometry, with
the nearest-neighbor spin spacing $d_{x}=d_{y}$. Due to the $C_{4v}$ point
group symmetry of the system, the corresponding 2D Zak phases are ($0,0$), ($%
\pm \pi ,\pm \pi $), ($2\pi ,2\pi $), while no such higher-order topological phases exist in the 2D SSH model.

\subsection{Edge states}

After discussing  the topological invariants, we are now in a position to study
topological edge states in the 2D spin-phononic system. To show the behavior of
edge states, here we consider a 2D strip structure with the periodic
boundary condition in the $y$ direction and $N_{x}$ unit cells in the $x$
direction \cite{wakabayashi2010electronic,wakabayashi2012nanoscale}. In this case, the Floquet engineering spin-spin interaction is
translationally invariant only along the $y$ direction.

\begin{figure}[tbp]
\includegraphics[width=8.6cm]{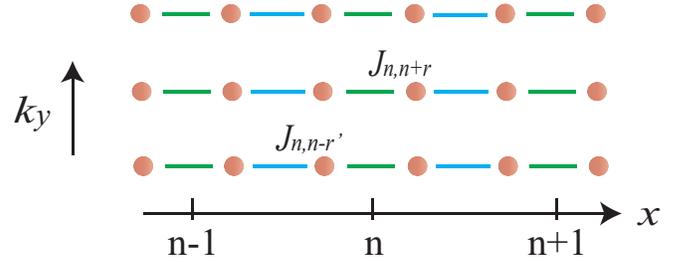}
\caption{(Color online) Schematic diagram of the 2D spin-spin strip structure. After the Fourier transformation, there are a set of 1D spin-spin interaction arrays indexed by a
continuous parameter $k_{y}$. For simplicity, only the nearest-neighbor interactions are illustrated.}
\label{fig_energy}
\end{figure}

As sketched in Fig. 8, after Fourier transformation in the $y$ direction, the two-dimensional
strip can be reduced to a set of 1D spin-spin interactions indexed by a
continuous parameter $k_{y}$. The
two-dimensional Hamiltonian can be rewritten as
\begin{eqnarray}
H_{2D}(k_{y}) &=&\underset{n,r,r^{\prime }}{\sum }\hbar \lbrack \mathcal{J}%
_{n,n+r}(A_{n}B_{n+r}^{\dagger }+C_{n}D_{n+r}^{\dagger } \notag \\
&&+A_{n}C_{n}^{\dagger }e^{-ik_{y}r}+B_{n}D_{n}^{\dagger }e^{-ik_{y}r}) \notag \\
&&+\mathcal{J}_{n,n-r^{\prime }}^{\prime }(A_{n}B_{n-r^{\prime }}^{\dagger
}+C_{n}D_{n-r^{\prime }}^{\dagger } \notag \\
&&+A_{n}C_{n}^{\dagger }e^{ik_{y}r^{\prime }}+B_{n}D_{n}^{\dagger
}e^{ik_{y}r^{\prime }})+H.c.],
\end{eqnarray}%
with $n=1,2,\ldots ,N_{x}$. In the single-excited state subspace
\begin{equation}
\psi (k_{y})=\underset{n}{\sum }(a_{n}A^{\dagger }_{n}+b_{n}B^{\dagger
}_{n}+c_{n}C^{\dagger }_{n}+d_{n}D^{\dagger }_{n})\left\vert 0\right\rangle,
\end{equation}%
where $%
a_{n},b_{n},c_{n},d_{n}$ denote the amplitudes of occupying probability in
the $n$th cell, respectively. Substituting Eqs. (57)-(58) to the
eigenvalue equation%
\begin{equation}
H_{2D}(k_{y})\psi (k_{y})=E\psi (k_{y}),
\end{equation}
we obtain the following set of equations of motion%
\begin{eqnarray}
Ea_{n} &=&f^{\ast }(k_{y})c_{n}+\mathcal{J}_{n,n+r}b_{n+r}+\mathcal{J}%
_{n,n-r^{\prime }}^{\prime }b_{n-r^{\prime }}, \notag \\
Eb_{n} &=&f^{\ast }(k_{y})d_{n}+\mathcal{J}_{n,n-r}a_{n-r}+\mathcal{J}%
_{n,n+r^{\prime }}^{\prime }a_{n+r^{\prime }}, \notag \\
Ec_{n} &=&f(k_{y})a_{n}+\mathcal{J}_{n,n+r}d_{n+r}+\mathcal{J}%
_{n,n-r^{\prime }}^{\prime }d_{n-r^{\prime }}, \notag \\
Ed_{n} &=&f(k_{y})b_{n}+\mathcal{J}_{n,n-r}c_{n-r}+\mathcal{J}%
_{n,n+r^{\prime }}^{\prime }c_{n+r^{\prime }}.
\end{eqnarray}
For the open boundaries $x=0,N_{x}+1$, the following amplitudes will be
vanished,
\begin{eqnarray}
b_{0} &=&d_{0}=0, \notag \\
a_{N_{x}+1} &=&c_{N_{x}+1}=0.
\end{eqnarray}%
In this way, we can analytically drive the edge modes.

In Figs. 7(b)-(d), we numerically calculate the resulting projected band structures with $%
N_{x}=11$. From the energy spectrum in the $k_{y}$ direction, we also verify the existence of edge states. The number of projected bands is determined by $%
N_{x}$. For the trivial case
with $a_{0}=4$, there exist only the bulk modes (blue curves), no gapless
modes emerge. While for the topological nontrivial case with $a_{0}=-11$ and
$21$, we see that the edge modes (red dash lines) appear inside the energy
band gaps. When $a_{0}=21$, the topological invariant winding number $%
\mathcal{W}=2$, and there are four zero-energy eigenstates, two of which are
degenerated. In addition, we can also notice that the energy spectrum are
symmetric with respect to the $E=0$, which is related to the chiral symmetry
of the system.

\section{robust quantum state transfer}

Topological nontrivial spin-spin interactions host zero-energy bound states at both ends. In the following, we show that the topological edge states can be employed as a quantum channel between distant qubits. Since quantum information can be transferred directly between the boundary spins, the intermediate spins are virtually excited during the process, which ensures the robust quantum state transfer \cite{lemonde2019quantum,mei2018robust}.

Taking into account the coupling of the system with the environment in the Markovian approximation, the evolution of the system follows the master equation
\begin{equation}
\dot{\rho}=-\frac{i}{\hbar }[H_{1D},\rho ]+\overset{2N}{\sum_{j=1}}\gamma
_{s}\mathcal{D}[\sigma _{j}^{z}]\rho ,
\end{equation}%
with $\sigma _{j}^{z}=|e\rangle _{j}\langle e|-|g\rangle _{j}\langle g|$, $%
\gamma _{s}$ the spin dephasing rate of the single SiV centers, and $\mathcal{D}[O]\rho
=O\rho O^{\dagger }-\frac{1}{2}\rho O^{\dagger }O-%
\frac{1}{2}O^{\dagger }O\rho $ for a given operator $O$.

To verify the theoretical results, we perform
numerical calculations by using the QuTiP library for the 1D spin array with $N=3$. Here we take the excited left end spin as the initial condition. As illustrated in Fig. $9(a)$, we obtain the significant Rabi oscillation of the left end spin. This implies that there are indeed quantum state transfer between the two ends of the spin array. However, for the non-topological condition, no
direct quantum state transfer can be seen, as shown in Fig. $9(c)$, since in the topological trivial regime, the eigenstates are the superposition of entire spin arrays. In addition,
we simulate the excitation dynamics for different parameters of the periodic driving. Compared Fig. $9(a)$
with Fig. $9(b)$, we see that the localization of the edge states is more obvious when setting $a_{0}=12$. While for the case with $a_{0}=24$, it takes shorter time for accomplishing
quantum state transfer. Finally,
we also consider the effect of spin dephasing on quantum state transfer.
As shown  in Fig. $9(a)$, when setting the dephashing rate $%
\gamma _{s}=1\times 10^{-4}J_{0}$, which is closed to the practical experimental conditions, the fidelity can reach $%
0.9$. The numerical results can be optimized by adjusting the parameters of the periodic driving field.

\begin{figure}[tbp]
\includegraphics[width=8.6cm]{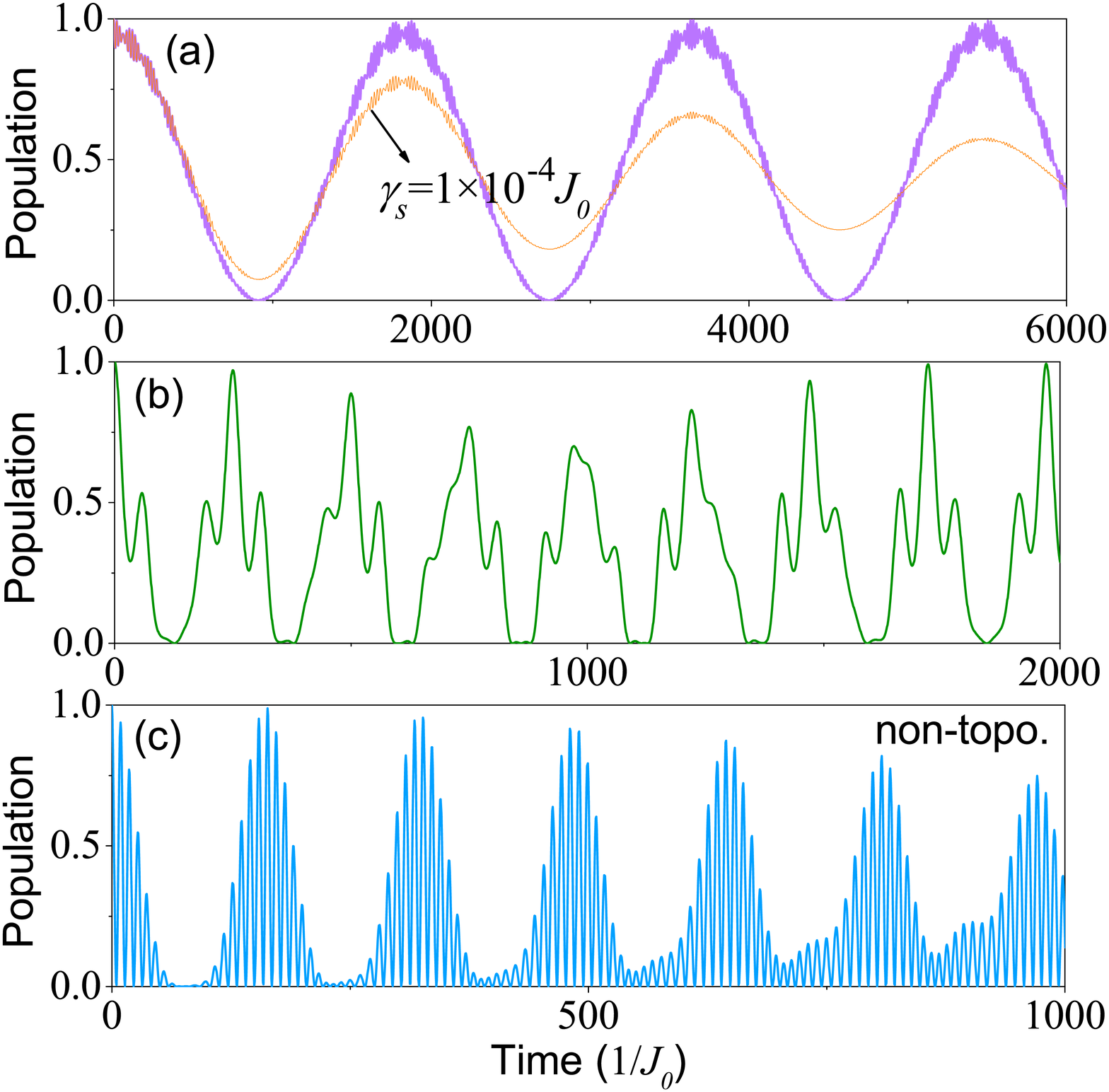}
\caption{(Color online) Excitation dynamics of the left end spin for various parameter settings of periodic driving: (a) $a_{0}=12$. (b) $a_{0}=24$. (c) $a_{0}=-2$. In $(a)$, we also add the result in the case of $%
\gamma _{s}=1\times 10^{-4}J_{0}$. Here we consider $N=3$. Other parameters are the same as those in Fig. $3$.}
\label{fig_energy}
\end{figure}

\section{Experimental feasibility}

We consider a 2D spin-phononic crystal network, where SiV centers are individually embedded in the nodes of a phononic crystal with square geometry. Based on state-of-art nanofabrication techniques, several experiments have demonstrated the generation of color center arrays through ion implantation \cite{toyli2010chip}. The fabrication of nanoscale mechanical structures with diamond crystals has been realized experimentally, as proposed in Refs. \cite{chan2012optimized,burek2016diamond,burek2017fiber}. Furthermore, owing to the advantage of the scalable nature of nanofabrication, the extension of phononic crystal structures to high dimensions is experimentally
feasible, and extensive research has been conducted \cite{pennec2010two,sukhovich2008negative,ding2019experimental,serra2018observation,he2016acoustic,safavi2010optomechanics,vasseur2001experimental,yang2004focusing}.

For the diamond phononic crystal illustrated in Fig. 1(a), the material properties are $E=1050$ GPa, $%
\nu =0.2$, and $\rho =3539$ kg/m$^{3}$. The lattice constant and cross
section of phononic crystal are $a=100$ nm and $A=100\times 20$ nm$^{2}$, and
the sizes of the elliptical holes are $(b,c)=(30,76)$ nm. With these carefully designed parameters, we derive a phononic band edge frequency $%
\omega _{BE}/2\pi =44.933$ GHz. The ground state transition frequency of SiV
center is about $46$ GHz, which is exactly located in a phononic band gap. The coupling between the SiV center and phononic crystal
mode $k$ is given by $g_{k}=\frac{d}{v_{l}}\sqrt{\frac{\hbar \omega _{BE}}{%
4\pi \rho aA}}\xi (\vec{r})$ \cite{sohn2018controlling}, where $d/2\pi \sim 1$ PHz is the strain
sensitivity, $v_{l}=1.71\times 10^{4}$ m/s is the speed of sound in diamond, and
$\xi (\vec{r})$ is the dimensionless strain distribution at the position of
the SiV center $\vec{r}$. Here we assign $\xi (\vec{r})=1$ \cite{safavi2010optomechanics}. Then, we can obtain the
effective SiV-phononic coupling rate as $g_{k}/2\pi \simeq 100$ MHz. In the large
detuning regime, $g\sim 0.1g_{k}$, the band gap engineered spin-phononic
coupling rate $g_{c}=g\sqrt{2\pi a/L_{c}}\simeq 2\pi \times 25$ MHz \cite{li2019band}.

In addition, we should consider the decoherence of the SiV-phononic crystal setup. For the SiV color center in diamond, at mK temperatures, the spin dephasing time is about $%
\gamma _{s}/2\pi =100$ Hz \cite{PhysRevLett.120.053603}. As for phononic crystals, the mechanical quality
factor is $Q\sim 10^{7}$, which can be achieved and further improved by
using 2D phononic crystal shields \cite{chan2012optimized}. In this case, we derive the mechanical
dampling rate $\gamma _{m}/2\pi =4.5$ kHz. As calculated above, the band gap
engineered spin-phononic coupling strength is $g_{c}/2\pi \simeq 25$ MHz,
which considerably exceeds both $\gamma _{s}$ and $\gamma _{m}$, resulting
in the strong strain interplay between the SiV centers and phonon crystal
modes. For the nearest neighbour spins with $d_{0}=a$, the bare
spin-spin interaction $J_{0}=\frac{g_{c}^{2}}{2\Delta _{BE}}%
\simeq 2\pi \times 4.1$ MHz. For the quantum state transfer in Fig. 9(a), the period is $%
\mathcal{T}=900/J_{0}\simeq 35$ $\mu$s, which is much shorter than the SiV spin coherence time ($%
T_{2}^{\ast }\sim 10$ ms) \cite{sukachev2017silicon}. Therefore, with the practical experimental conditions, this proposal can be implemented to achieve high-fidelity quantum state transfer.

\section{conclusion}

To conclude, we explore the topological quantum properties in two-dimensional SiV-phononic crystal networks. Applying a special periodic drive to the SiV centers, the phononic band-gap mediated spin-spin interactions exhibit a topologically protected chiral symmetry. Then, we study the topological properties of  the 1D and 2D Floquet engineering SiV center arrays, respectively. For the periodic driving with suitably chosen parameters, we analyse and simulate the corresponding topological invariants. We show that, under the appropriate driving fields,  higher-order topological phases can be simulated in the spin-phononic crystal structures.

In contrast to the SSH model, the present Floquet engineering spin-spin interaction can be selectively controlled by modulating the periodic driving, which is essential for generating the necessary symmetries of the topological protection. More interestingly, we present rich topological Zak phases in this work. Owing to the highly controllable and tunable nature of the periodic driving, it is feasible to investigate the topological properties of the trimer case in SiV-phononic crystal systems. As an outlook, this proposal can be explored to study chiral quantum acoustics, topological quantum computing, and the implementation of hybrid quantum networks.
\section*{Acknowledgments}

This work is supported by the National Natural Science Foundation of
China under Grant No.
11774285, and Natural Science Basic Research Program of Shaanxi (Program No.
2020JC-02).

%

\end{document}